\documentclass[letterpaper]{IEEEtran}
\ifCLASSINFOpdf
\else
\fi
\usepackage{amsthm,amsmath,amssymb,mathtools,bm,etoolbox,float,hyperref}

\usepackage{graphicx,float,caption}
\usepackage{tikz,float,cite}
  \newlength\fheight
\newlength\fwidth
\usepackage{tkz-euclide}
\def\BibTeX{{\rm B\kern-.05em{\sc i\kern-.025em b}\kern-.08em
    T\kern-.1667em\lower.7ex\hbox{E}\kern-.125emX}}
    \usepackage{pgfplots} 
\usepackage{pgfgantt}
\usepackage{pdflscape}
\pgfplotsset{compat=newest} 
\pgfplotsset{plot coordinates/math parser=false}
\pgfplotsset{every  tick/.style={black,},ylabel style={font=\tiny},xlabel style={font=\tiny},tick label style={font=\tiny},legend style= {font=\scriptsize},
minor x tick num=1,minor y tick num=1,xminorticks=true,yminorticks=true,}
\begin{document}

\title{Adaptive Self-Interference Cancellation for Full-Duplex Wireless Communication Systems
}

\author{Elyes~Balti,~\IEEEmembership{Student~Member,~IEEE,}
        and~Brian~L.~Evans,~\IEEEmembership{Fellow,~IEEE}% <-this % stops a space
\thanks{Elyes Balti and Brian L. Evans are with the Wireless Networking and Communications Group, Department of Electrical and Computer Engineering, The University
of Texas at Austin, Austin, TX 78712 USA (e-mails: ebalti@utexas.edu, bevans@ece.utexas.edu).}% <-this % stops a space
}

\maketitle

\begin{abstract}
In this letter, we consider single-cell, single-user systems wherein uplink and downlink user equipment communicate with a full-duplex relay. Due to the near-far problem, the self-interference (SI) can be 100-1000x the received signal power. In this context, we consider the
adaptive Least Mean Squares (LMS) algorithm to estimate the SI channel and then subtract the SI from the desired received signal before the analog-to-digital converter (ADC). We measure the robustness of this technique in terms of bit error rate (BER) and spectral efficiency.
\end{abstract}


\IEEEpeerreviewmaketitle

\section{Introduction}
\IEEEPARstart{F}{ull-Duplex} (FD) systems have recently gained enormous attention in academia and industry due to its potential to reduce latency and double spectral efficiency in the link budget compared to the half-duplex relays that transmit and receive in different time slots \cite{thesis}. Because of these benefits, FD systems can be integrated in applications requiring high data rate, such as platooning, autonomous driving, and vehicular clouds \cite{surv,3gpp2,3gpp3,3gpp4,3gpp5,3gpp6}, Since it transmits and receives at the same resource blocks, FD transceivers are vulnerable to the near-far problem. In fact, the FD receive antenna receives a signal from its transmit antenna side that can be 100-1000x stronger than the desired received signal due to the propagation losses over large distances \cite{masmoudi}. In other terms, the near-far problem is translated into a loop-back self-interference (SI) that can saturates the analog-to-digital converter (ADC) resulting in severe degradation of the reliability performance. Therefore, SI cancellation techniques are necessary to eliminate the loop-back SI and reduce it below the noise floor. 

Conventional SI cancellation methods have been discussed on the literature and they are mainly classified into hardware and systems based techniques. Hardware based approach are mainly antenna separation, isolation, polarization \cite{2,3,4}, directional antennas \cite{5,6,7} or antennas placement to create null space at the receive arrays which achieves a 15 dB of SI reduction \cite{9,masmoudi}. While, system based approaches are the beamforming techniques, i.e., designing the analog phases shifters to eliminate the SI before the ADC and/or designing the baseband receiver, which is the last line of defense, to remove the SI \cite{zf,adaptive}.

In this work, we consider a single-user and single-cell system wherein the uplink and downlink users equipments (UEs) are equipped with single antenna and communicate through the FD relay. In the first stage, we consider the adaptive Least Mean Square (LMS) algorithm to estimate the SI channel while the estimated SI signal is subtracted from the desired uplink signal at the receive antenna of the FD relay in the second stage. We consider the multicarrier Orthogonal Frequency Division Multiplexing (OFDM) transmission mode that is used for the training pilots for the estimation stage as well as the payload data. The performances are measured in terms of bit error rate (BER) and spectral efficiency.

The remainder of the paper is structured as follows: Section II describes the signals and channels models while the SI cancellation method is discussed in Section III. Section IV provides the numerical results following their discussions while the concluding remarks are reported in Section V.

\begin{figure}[t]
    \centering
    \includegraphics[width=\linewidth]{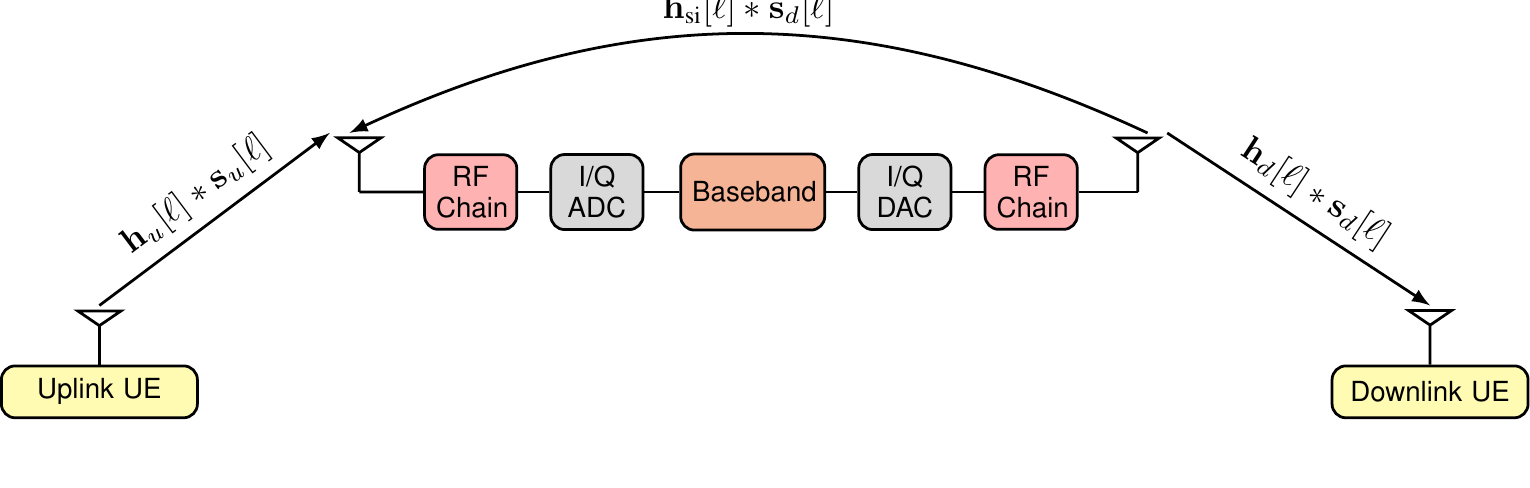}
    \caption{Dual-hop FD relaying system. The uplink and downlink UEs communicates with the FD relay. Note that the uplink and downlink data of the UEs are independent.}
    \label{system}
\end{figure}

\section{System Model}
We consider a single-cell and single-user scenario wherein the uplink and downlink users equipments (UEs) communicate with FD relay as illustrated by Fig.~\ref{system}.

The uplink received signal at the discrete time instant $\ell$ is given by
\begin{equation}
    \bold{y}_u[\ell] = \bold{h}_u[\ell]*\bold{s}_u[\ell] + \bold{h}_{\text{si}}[\ell]*\bold{s}_d[\ell] + \bold{n}[\ell]
\end{equation}
where $(*)$ is the convolution operator, $\bold{s}_u[\ell]$ and $\bold{s}_d[\ell]$ are the uplink and downlink sequences, $\bold{n}[\ell]$ is the Additive White Gaussian Noise (AWGN) at the relay, while $\bold{h}_u[\ell]$ and $\bold{h}_{\text{si}}[\ell]$ are the uplink and SI channels, respectively. Note that $\bold{h}_{\text{si}}[\ell]$ consists of line-of-sight (LOS) or also termed as the direct or internal path and non-line-of-sight (NLOS) paths caused by the external reflections. A general model of the SI channel is given by
\begin{equation}
\bold{h}_{\text{si}}[\ell] = \sqrt{\frac{\kappa}{\kappa + 1}} \bold{h}_{\text{los}}[\ell] + \sqrt{\frac{1}{\kappa + 1}} \bold{h}_{\text{nlos}}[\ell]   
\end{equation}
where $\kappa$ is the Rician factor. Note that $\bold{h}_{\text{los}}[\ell]$ is static and depends on the geometry of the FD transceiver while $\bold{h}_{\text{nlos}}[\ell]$ is generally probabilistic to model the external random reflections.

For downlink scenario, the FD relay transmits the sequences $\bold{s}_d[\ell]$ to the downlink UE which is interference-free. The downlink received signal is expressed by
\begin{equation}
\bold{y}_d[\ell] = \bold{h}_d[\ell]*\bold{s}_d[\ell] + \bold{n}[\ell]    
\end{equation}

\section{Adaptive Self-Interference Cancellation}
\subsection{Training and Estimation}
In this stage, the FD relay exchanges the training pilots through the internal and external paths. During this stage, we apply the steepest descent to construct the coefficients of the Finite Impulse Response (FIR) filter taps which is the estimate the SI channel. The training frame contains an OFDM symbol (cyclic prefix and pilots) generated from QPSK symbols. The output of the training stage is the SI channel estimate $\widehat{\bold{h}}_{\text{si}}$.

\subsection{Brief Review}
We consider a cost function $\mathcal{I}(\bold{w})$ that is a \textit{continuously differentiable function} of some unknown weight vector $\bold{w}$. This function maps the vector elements of $\bold{w}$ into real numbers. We aim to obtain an optimal solution $\bold{w}_0$ that satisfies the following condition
\begin{equation}
\mathcal{I}(\bold{w}_0) \leq \mathcal{I}(\bold{w})~~\text{for all}~\bold{w} 
\end{equation}
We start with initial guess denoted by $\bold{w}(0)$ and then we generate a sequence of weight vectors $\bold{w}(1),\bold{w}(2),\ldots$ such that the cost function $\mathcal{I}(\bold{w})$ is decreased at each adaptation cycle such as
\begin{equation}
\mathcal{I}(\bold{w}(\ell+1)) \leq \mathcal{I}(\bold{w}(\ell))    
\end{equation}
where $\bold{w}(\ell)$ is the old value of the weight vector and $\bold{w}(\ell+1)$ is the updated version. To converge to the optimal solution, the successive adjustments applied to the weight vector $\bold{w}$ are in the direction of steepest descent, i.e., in a direction opposite to the gradient vector of the cost function $\mathcal{I}(\bold{w})$, which is denoted by $\nabla \mathcal{I}(\bold{w}) = \partial \mathcal{I}(\bold{w})/\partial \bold{w}$.

Accordingly, the steepest descent is described as
\begin{equation}
\bold{w}(\ell+1) = \bold{w}(\ell) - \frac{1}{2}\mu \frac{\partial \mathcal{I}(\bold{w}(\ell))}{\partial \bold{w}(\ell)} 
\end{equation}
where $\mu$ is the step-size.

We further define the input training sequences at time $\ell$ $\bold{u}(\ell) = \left[u(\ell),u(\ell-1),\ldots,u(\ell-L+1)  \right]^{\text{T}}$, $L$ is the number of the filter taps and $(\cdot)^{\text{T}}$ is the Transpose operator. At the $\ell$-th adaptation cycle, we produce an estimation error as 
\begin{equation}
e(\ell) = d(\ell) - \bold{w}^{\text{H}}(\ell)\bold{u}(\ell)    
\end{equation}
where $(\cdot)^{\text{H}}$ is the Hermitian operator, $e(\ell)$ and $d(\ell)$ are the estimation error and the desired signal at the time index $\ell$, respectively. If the training sequences $\bold{u}(\ell)$ and the desired signal $d(\ell)$ are jointly stationary, then the mean square error $\mathcal{I}(\bold{w}(\ell))$ at time index $\ell$ is a quadratic function of the tap-weight vector $\bold{w}$. According to \cite[Eq.~(2.50)]{book}, the mean square error is expressed by
\begin{equation}
\mathcal{I}(\bold{w}(\ell)) = \sigma_d^2 - \bold{w}^{\text{H}}(\ell)\bold{p} -  \bold{p}^\text{H}\bold{w}(\ell) +  \bold{w}^{\text{H}}(\ell)\bold{R}\bold{w}(\ell)  
\end{equation}
where $\sigma_d^2$ is the variance of the desired signal, $\bold{p}$ is the cross-correlation vector between $\bold{u}(\ell)$ and $d(\ell)$ and $\bold{R}$ is the correlation matrix of the input training $\bold{u}(\ell)$. Note that the gradient vector is obtained by
\begin{equation}
\nabla   \mathcal{I}(\bold{w}(\ell)) = -2\bold{p} + 2\bold{R} \bold{w}(\ell)  
\end{equation}
Therefore, the estimate of the output weight vector is expressed by
\begin{equation}
 \bold{w}(\ell+1) =  \bold{w}(\ell) + \mu \left[ \bold{p} - \bold{R} \bold{w}(\ell) \right]     
\end{equation}
which describes the mathematical formulation of the steepest descent algorithm for Wiener filtering.

A necessary and sufficient condition for the convergence or stability of the steepest descent is that the step size $\mu$ has to satisfy the double inequality
\begin{equation}
0 < \mu < \frac{2}{\sigma_{\text{max}}}    
\end{equation}
where $\sigma_{\text{max}}$ is the largest eigenvalue of the correlation matrix $\bold{R}$. In the sequel, we express the step size as 
\begin{equation}\label{alpha}
\mu = \frac{\alpha}{\sigma_{\text{max}} }   
\end{equation}
Regarding our framework, the output weight vector $\bold{w}$ is nothing but the estimate of the SI channel $\widehat{\bold{h}}_{\text{si}}$.

Adaptive LMS has low complexity compared to other algorithms. For a finite impulse response filter of $L$ taps, the steepest descent requires $2L + 2$ FLOPS/update while Recursive Least Square requires $L^2$ FLOPS/update.
\subsection{Interference Cancellation}
Once the SI channel is estimated in the first stage, we subtract the estimated SI signal from the uplink received signal. The uplink received signal before the ADC and after the interference cancellation is expressed as
\begin{equation}\label{uplink}
\bold{y}_{u}[\ell] = \bold{h}_u[\ell]*\bold{s}_u[\ell] + \underbrace{\left(\bold{h}_{\text{si}}[\ell] - \widehat{\bold{h}}_{\text{si}}[\ell]\right) *\bold{s}_d[\ell]}_{{\scriptsize{\textsf{Residual self-interference}}}} + \bold{n}[\ell]
\end{equation}

\section{Numerical Results}
We assume that the uplink $\bold{h}_{u}[\ell]$ and NLOS $\bold{h}_{\text{nlos}}[\ell]$ channels are Complex Gaussian distributed while the internal LOS $\bold{h}_{\text{los}}[\ell]$ channel is a $\textsf{sinc}(\cdot)$ pulse. System parameters and their values (unless otherwise stated) are summarized in Table \ref{param}.
\begin{table}[H]
\caption{System Parameters}
\label{param}
\centering
\begin{tabular}{|r|c|l|}
\hline
\bfseries Subsystem & \bfseries Parameter & \bfseries Value\\
\hline
Channel      & No.\ Channel Taps ($L$) & 10\\
             & Rician Factor& 5 dB\\
\hline
OFDM         & Cyclic Prefix & $L-1$\\
Modulation/  & Bits per OFDM Symbol & 32400\\
Demodulation & Subcarrier Modulation & QPSK \\
             & Frames (OFDM Symbols) & $10^5$\\
\hline
Error        & Channel Coding & LDPC\\
Correction   & Coding Rate & 1/2\\
\hline
Data Converters  & ADC/DAC Resolution & $\infty$ \\
\hline
Self-Interference & Signal-to-Interference Ratio (SIR) & -60 dB \\
\hline
Self-Interference & Iterations & 100\\
Cancellation      & $\alpha$ in (\ref{alpha}) & 0.125 \\
\hline
\end{tabular}
\end{table}
During training, we generate a pilot frame (OFDM symbol plus cyclic prefix) containing 31400 LDPC coded bits mapped into QPSK subcarrier amplitudes, which are then combined via the fast Fourier transform to an OFDM symbol. The pilot frame is input to the steepest descent method for 100 iterations to estimate the SI channel. Then, we randomly generate another 31400 LDPC coded bits mapped to an OFDM symbol to construct the data frame for the uplink and downlink UEs. To cancel the SI, the downlink data frame is convolved with the estimated SI channel and then subtracted from the uplink received signal per (\ref{uplink}). In simulation, we send $10^{5}$ (Monte Carlo) frames for training, uplink and downlink data, and then average the BER and spectral efficiency.

\begin{figure}[t]
\centering
\setlength\fheight{5.5cm}
\setlength\fwidth{7.5cm}
\definecolor{mycolor1}{rgb}{0.00000,0.44700,0.74100}%
\definecolor{mycolor2}{rgb}{0.85000,0.32500,0.09800}%
\definecolor{mycolor3}{rgb}{0.92900,0.69400,0.12500}%
\begin{tikzpicture}

\begin{axis}[%
width=0.951\fwidth,
height=\fheight,
at={(0\fwidth,0\fheight)},
scale only axis,
xmin=-10,
xmax=30,
xlabel style={font=\color{white!15!black}},
xlabel={\textsf{SNR (dB)}},
ymode=log,
ymin=0.0001,
ymax=1,
yminorticks=true,
ylabel style={font=\color{white!15!black}},
ylabel={ \textsf{BER}},
axis background/.style={fill=white},
axis x line*=bottom,
axis y line*=left,
legend style={at={(0.03,0.03)}, anchor=south west, legend cell align=left, align=left, draw=none,fill=none}
]

\draw[<->, >=latex,line width=1pt] (29.5,0.000113764109347443)-- node[below,sloped] {} node[above,sloped]{\scriptsize{\sffamily{Error Improvement} } }(29.5,      0.246943827160494); 

\addplot [color=blue,dash pattern={on 10pt off 1pt on 1pt off 1pt}, line width=1.3pt]
  table[row sep=crcr]{%
-10	0.49892046815874\\
-8	0.450089891152907\\
-6	0.394241067437697\\
-4	0.333130616394622\\
-2	0.270011954773346\\
0	0.209209442009609\\
2	0.154978188642263\\
4	0.110195594366273\\
6	0.0757126913223411\\
8	0.0506558108969115\\
10	0.0332410076817197\\
12	0.0215209719576469\\
14	0.0138071984977693\\
16	0.00880547820909215\\
18	0.00559392663990164\\
20	0.00354486413572203\\
22	0.00224281126359558\\
24	0.00141758010876688\\
26	0.000895415762602167\\
28	0.000565361343700586\\
30	0.000356875223019145\\
};
%\addlegendentry{Uncoded (Interference-Free)}
\addlegendentry{\scriptsize{\textsf{Uncoded (Interference-Free)}}}

\addplot [color=blue, line width=1.3pt]
  table[row sep=crcr]{%
-10	0.293431660714286\\
-8	0.276616512566138\\
-6	0.255828127425044\\
-4	0.229944155202822\\
-2	0.199528981481481\\
0	0.164198880731922\\
2	0.124047856261023\\
4	0.0805351406525573\\
6	0.0427706031746032\\
8	0.0194989781746032\\
10	0.00843018650793651\\
12	0.00392938580246914\\
14	0.00208869356261023\\
16	0.00131918738977072\\
18	0.00091382385361552\\
20	0.000635802469135802\\
22	0.000447492724867725\\
24	0.000309039682539682\\
26	0.00020198897707231\\
28	0.000133043430335097\\
30	0.000113764109347443\\
};
%\addlegendentry{LDPC (After Cancellation)}
\addlegendentry{\scriptsize{\textsf{LDPC (After SI Cancellation)}}}

\addplot [color=blue, dash pattern={on 10pt off 1pt on 0pt off 0pt},line width=1.3pt]
  table[row sep=crcr]{%
-10	0.205299520061728\\
-8	0.193690038580247\\
-6	0.179034970679012\\
-4	0.161238403549383\\
-2	0.139540085648148\\
0	0.115093857253086\\
2	0.0868172654320988\\
4	0.0565508804012346\\
6	0.0298383449074074\\
8	0.0136724915123457\\
10	0.00575559799382716\\
12	0.00261361342592593\\
14	0.00142823456790123\\
16	0.000919013117283951\\
18	0.000597538580246914\\
20	0.000411276234567901\\
22	0.000278281635802469\\
24	0.000189248456790123\\
26	0.000113222222222222\\
28	8.31913580246914e-05\\
30	4.16550925925926e-05\\
};
%\addlegendentry{LDPC (Interference-Free)}
\addlegendentry{\scriptsize{\textsf{LDPC (Interference-Free)}}}

\addplot [color=blue, dotted,line width=1.3pt]
  table[row sep=crcr]{%
-10 0.247026697530864\\
-8 0.247265277777778\\
-6 0.247042283950617\\
-4 0.247146296296296\\
-2 0.247238425925926\\
0 0.247097376543210\\
2 0.246609104938272\\
4 0.247040277777778\\
6 0.246940277777778\\
8 0.246800308641975\\
10 0.246798919753086\\
12 0.246595061728395\\
14 0.246822376543210\\
16 0.246789506172840\\
18 0.246548302469136\\
20 0.246610339506173\\
22 0.246760185185185\\
24 0.246764351851852\\
26 0.247010030864198\\
28 0.246795679012346\\
30 0.246943827160494\\
};
\addlegendentry{\scriptsize{\textsf{LDPC (With Interference)}}}
\end{axis}
\end{tikzpicture}%
    \caption{Bit error rate (BER) vs. SNR.
    In the presence of self-interference (SI) with SIR of -60 dB,
    the LDPC BER is flat because the received uplink signal is
    swamped by interference. Compare this to LDPC after SI cancellation
    which follows the LDPC curve when there is no interference with a small
    SNR gap. Uncoded BER is also shown when there is no interference.}
    \label{ber}
\end{figure}
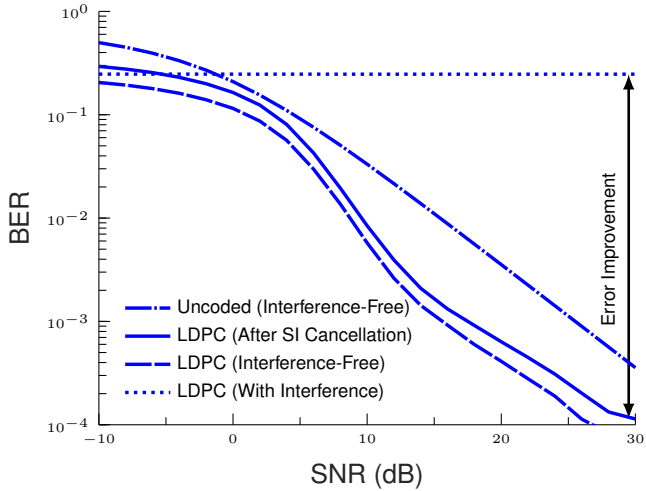

Without SI cancellation, as shown in Fig.\ \ref{ber}, coded BER is very high (around 0.25) and stays approximately constant with increasing SNR due to the SIR of -60 dB. BER is dramatically reduced after SI has been canceled by the adaptive LMS algorithm. The proposed technique achieves BER of $10^{-4}$ around 30 dB of SNR. Although the proposed technique achieves a high amount of SI reduction, residual SI remains which is visible as a small gap with the BER implemented for interference-free (downlink UE) and LDPC coding.

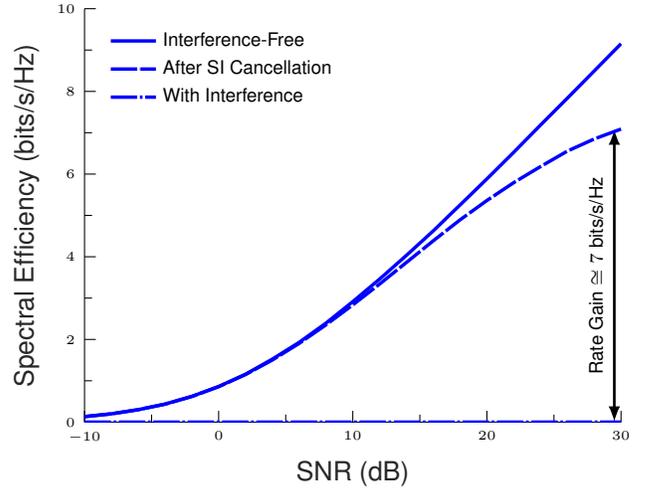
\begin{figure}[t]
\centering
\setlength\fheight{5.5cm}
\setlength\fwidth{7.5cm}
\definecolor{mycolor1}{rgb}{0.00000,0.44700,0.74100}%
\definecolor{mycolor2}{rgb}{0.85000,0.32500,0.09800}%
\definecolor{mycolor3}{rgb}{0.92900,0.69400,0.12500}%
\begin{tikzpicture}

\begin{axis}[%
width=0.951\fwidth,
height=\fheight,
at={(0\fwidth,0\fheight)},
scale only axis,
xmin=-10,
xmax=30,
xlabel style={font=\color{white!15!black}},
xlabel={\textsf{SNR (dB)}},
ymin=0,
ymax=10,
ylabel style={font=\color{white!15!black}},
ylabel={\textsf{Spectral Efficiency (bits/s/Hz)}},
axis background/.style={fill=white},
axis x line*=bottom,
axis y line*=left,
legend style={at={(0.03,0.97)}, anchor=north west, legend cell align=left, align=left, draw=none}
]

\draw[<->, >=latex,line width=1pt] (29.5,0.01)-- node[below,sloped] {} node[above,sloped]{\scriptsize{\sffamily{Rate Gain} $\cong 7$ \textsf{bits/s/Hz}}} (29.5,7.08);

\addplot [color=blue, line width=1.3pt]
  table[row sep=crcr]{%
-10	0.131719993990853\\
-8	0.199970821798769\\
-6	0.300686804287084\\
-4	0.435039848101375\\
-2	0.622546752679054\\
0	0.861962310692254\\
2	1.157357824044204\\
4	1.523095402195569\\
6	1.928469377976994\\
8	2.393746646488601\\
10	2.913848420953468\\
12	3.459601007988381\\
14	4.033684909617778\\
16	4.625871776272797\\
18	5.253269750151012\\
20	5.882713541640349\\
22	6.526703947306197\\
24	7.186024510189242\\
26	7.836601667039482\\
28	8.493796788980621\\
30	9.152748434204307\\
};
\addlegendentry{\scriptsize{\textsf{Interference-Free}}}

\addplot [color=blue,dash pattern={on 10pt off 1pt on 0pt off 0pt}, line width=1.3pt]
  table[row sep=crcr]{%
-10	0.133551894524288\\
-8	0.199923384839641\\
-6	0.300183787353708\\
-4	0.434362719674599\\
-2	0.619349871408430\\
0	0.856593301278925\\
2	1.149957870531349\\
4	1.501554757247806\\
6	1.903261948038450\\
8	2.352796544144154\\
10	2.834879637221770\\
12	3.339484868127765\\
14	3.862571776837691\\
16	4.385981939648533\\
18	4.892935863013173\\
20	5.360997731799671\\
22	5.792711069125762\\
24	6.181978210585982\\
26	6.555516500854586\\
28	6.848963593039300\\
30	7.089136719372055\\
};

\addlegendentry{\scriptsize{\textsf{After SI Cancellation}}}
\addplot [color=blue,dash pattern={on 10pt off 1pt on 1pt off 1pt}, line width=1.3pt]
  table[row sep=crcr]{%
-10	5.030281739310769e-05\\
-8	5.340570431490700e-05\\
-6	5.568613989271134e-05\\
-4	5.722952355666280e-05\\
-2	5.784794564466832e-05\\
0	6.360010951576200e-05\\
2	6.337994611764612e-05\\
4	6.496825025459327e-05\\
6	6.089048162402438e-05\\
8	6.333056342906681e-05\\
10	6.626819901329271e-05\\
12	6.505732153172508e-05\\
14	6.949537460304860e-05\\
16	6.739906122474401e-05\\
18	6.661345069930968e-05\\
20	6.224350313117632e-05\\
22	6.697591565661859e-05\\
24	6.435201320198251e-05\\
26	6.809532469672140e-05\\
28	6.473489998127334e-05\\
30	6.372470638100840e-05\\
};
\addlegendentry{\scriptsize{\textsf{With Interference}}}

\end{axis}
\end{tikzpicture}%
    \caption{Spectral efficiency results vs. SNR. 
    The proposed self-interference (SI) cancellation technique shows significant improvement in the
    presence of SI at an SIR of -60 dB, and 
    tracks the interference-free scenario which serves as a benchmark.}
    \label{rate}
\end{figure}
Fig.~\ref{rate} illustrates variations in spectral efficiency vs. SNR. As mentioned earlier, spectral efficiency is severely degraded by SI to be around 0.1 bits/s/Hz. We also observe the efficiency of the proposed technique to estimate and cancel the SI, and it compensates for roughly 7 bits/s/Hz of rate loss.

\section{Conclusion}
We considered a single-cell single user scenario wherein two UEs are communicating with a FD relay. We use an adaptive LMS method to cancel the SI which corrupts the received uplink UE signal. This technique achieved a high amount of SI reduction although some residual SI remained. We plan to extend this approach to support MIMO-OFDM systems in multiuser cellular case with low-resolution ADCs.\\
\textbf{\textit{Reproducible research:}} \href{https://github.com/ebalti/Full-Duplex-Steepest-Descent}{https://github.com/ebalti/Full-Duplex-Steepest-Descent}

\bibliographystyle{IEEEtran}
\bibliography{main.bib}
\end{document}